\documentclass[a4paper]{jpconf}
\usepackage{graphicx}
\usepackage{epsfig}
\usepackage{amssymb}
\DeclareGraphicsExtensions{.eps,.jpeg,.png}

\begin{document}
\title{Suppression of the slow component of BaF$_{2}$ crystal luminescence with a thin multilayer filter}

\author{AM~Artikov$^{1}$, V~Baranov$^{1}$, JA~Budagov$^{1}$, AN~Chivanov$^{2}$, YuI~Davydov$^{1}$, EN~Eliseev$^{3}$, EA~Garibin$^{4}$, VV~Glagolev$^{1}$, AV~Mihailov$^{3}$, PA Rodnyi$^{5}$, VV~Terechschenko$^{1}$ and II~Vasilyev$^{1}$}

\address{$^{1}$ JINR, Dubna, Russia}
\address{$^{2}$ “Ural-GOI”, Branch of JSC «PA «UOMP», St. Petersburg, Russia}
\address{$^{3}$ LLC “Optech”, St. Petersburg, Russia}
\address{$^{4}$ INCROM ltd, St. Petersburg, Russia}
\address{$^{5}$ St. Petersburg State Polytechnic University, St. Petersburg, Russia}

\ead{davydov@jinr.ru}

\begin{abstract}
The fast component of the barium fluoride (BaF$_{2}$) crystal luminescence with the emission peak at 220 nm allows those crystals to be employed in fast calorimeters operating in harsh radiation environment. However, the slow component with the emission peak at 330 nm and about 85\% of the total emission light could create big problems when working at a high radiation rate.
	In this work we report results of tests of multilayer filters that can suppress luminescence in the range from 250 nm to 400 nm, which covers most of the BaF$_{2}$ slow component luminescence. The filters are made by spraying layers of rare earth oxides on a quartz glass substrate. Filters typically comprise 200-220 layers.
	A few filters were prepared by spraying thin layers on quartz glass. The filters have a peak transmittance of about 70-80\% in the range of 200-250 nm. Measurements of the light output of the BaF$_{2}$ crystal with and without a filter between the crystal readout end and the PMT demonstrate substancial suppression of the slow component. To our knowledge, this kind of filters are produced and tested for the first time.
\end{abstract}

\section{Introduction}

The Mu2e experiment \cite{Mu2e_TDR} plans to search for neutrino less conversion of a muon into an electron in the field of a nucleus prohibited in the Standard Model. The goal of the experiment is the single event sensitivity with an aluminum target at a level of 2.5$\times$10$^{-17}$. Evolution of the experiment to Mu2e-II \cite{Mu2e-II} will lead to improvement of the single event sensitivity by an order of magnitude. That will require replacement of CsI crystals with BaF$_{2}$ crystals in the Mu2e electromagnetic calorimeter (EMC) .

Barium fluoride is an excellent candidate for the use in the EMC in harsh radiation environment due to its fast component of luminescence with the emission peak at 220 nm. However, the slow component of luminescence with the emission peak at 330 nm and about 85\% of the total emission light needs to be suppressed to use at high radiation rate.

The slow component of the BaF$_{2}$ crystal luminescence could be suppressed by doping BaF$_{2}$ with rare earth \cite{BaF2_suppress}, depositing an atomic layer of interference filter to get solar blind windows on the sensors \cite{Hitlin}, applying nanoparticle coatings on sensors, and using external interference optical filters.
In this work we report the results of the study of thin interference optical multilayer filters made by spraying layers of rare earth oxides on the quartz glass substrate. Such filters can suppress luminescence in the range of about 250 nm to 400 nm, which covers most of the BaF$_{2}$ slow component.

\section{Thin multilayer filters}

Calculation of the filter design, selection of film-forming materials, and complex analysis of sprayed filters were carried out by the specially developed program. Filters are formed by spraying thin layers of rare earth oxides on the substrate. Layers are made by electron-beam evaporation of selected materials.
Typically, filters comprise up to 200-220 layers depending on the optical range and suppression level.

A few samples of multilayer filters sprayed on the quartz glass  substrates (KU-2 type) were prepared for the tests. Quartz glass substrates are 30 mm in diameter and 3 mm thick. Quartz glass is an optimal material for the multilayer filter evaporation.

\begin{figure}[h]
\centering
\includegraphics[width=0.8\textwidth]{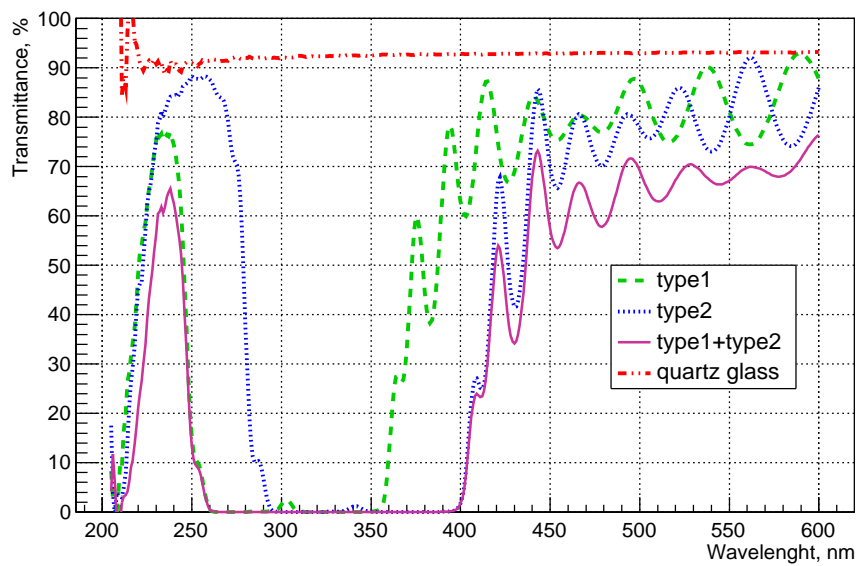}
\caption{Transmittances of filters: type1, type2, double type1+type2 filter. Transmittance of quartz glass itself is shown as well.}
\label{Filter_transmit}
\end{figure}

At the present stage it was possible to produce a filter with a suppression width of about 100 nm only. In order to provide suppression in the range of 250-400 nm, two different filters were produced with the suppression width between 250-350 nm and 300-400 nm, which we called "type1" and "type2" respectively.

The filter transmittances were measured with the Shimadzu SolidSpec-3700 DUV photo spectrometer. Figure~\ref{Filter_transmit} demonstrates transmittances of type1 and type2 filters separately and a type1+type2 pair. Transmittance of quartz glass of the same size, but with no evaporated filter is shown in Figure~\ref{Filter_transmit} as well. One can see that the pair of type1+type2 filters together provide suppression between 250 and 400 nm. However, transmittance of a double filter drops in the area 200-250 nm, which causes suppression of the fast component as well.

\section{Test results}

\begin{figure}[h]
\centering
\includegraphics[width=0.8\textwidth]{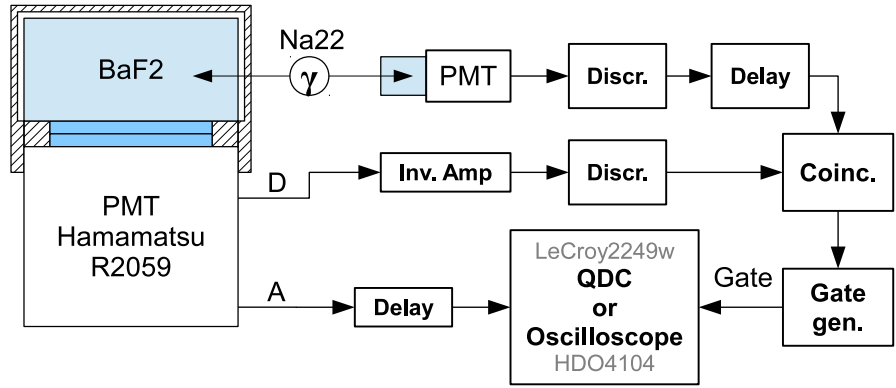}
\caption{The block diagram of the test setup.}
\label{setup}
\end{figure}

Filters were tested with a BaF$_{2}$ crystal with the diameter of 38 mm and height of 18 mm. All crystal surfaces were optically polished. The block diagram of the test setup is shown in Figure~\ref{setup}. The BaF$_{2}$ crystal was attached to the Hamamatsu R2059 photomultiplier (PMT) directly or with a pair of filters (type1+type2). No optical grease was used between the PMT, filters, and crystal. The crystal was completely wrapped with Tyvek paper except for the 30-mm-diameter window.

The BaF$_{2}$ crystal was irradiated with a $^{22}$Na gamma source. Triggers were provided by coincidence of signals from the last dynode of the R2059 PMT and another trigger counter with a smaller crystal (1 cm$^{3}$) when two back-to-back emitted 511 keV gammas hit both crystals. The LeCroy 2249W QDC or HDO4104 digitizing oscilloscope was employed for measurement of the anode signal from the R2059 PMT. The trigger signal produced the gate/start signal for the QDC/oscilloscope.

\begin{figure}[h]
\begin{minipage}{18pc}
\includegraphics[width=18pc]{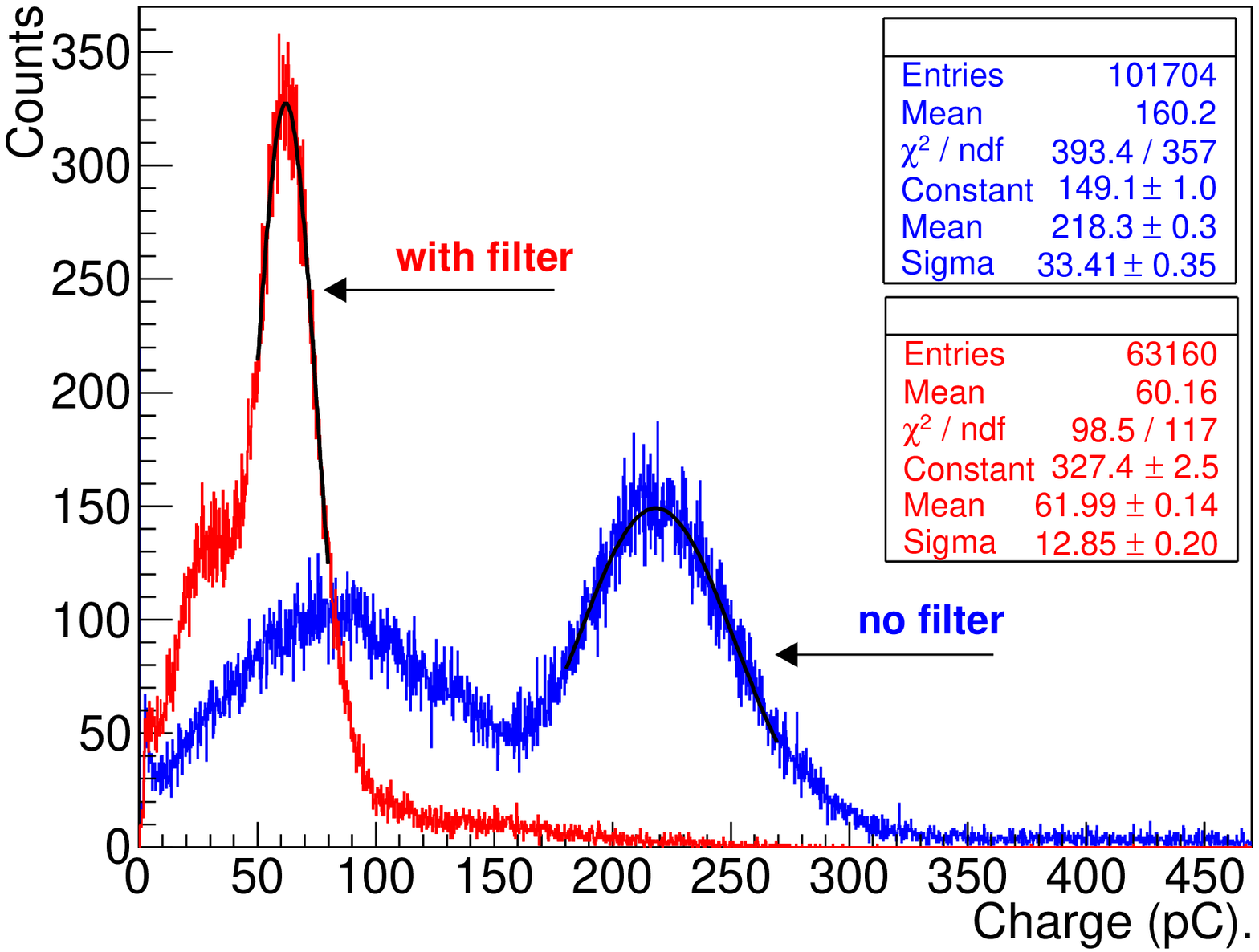}
\caption{\label{spectra_QDC}Spectra from the BaF$_{2}$ crystal irradiated with the $^{22}$Na gamma source without and with a filter. The data are taken with the 2249W QDC.}
\end{minipage}\hspace{2pc}%
\begin{minipage}{20pc}
\includegraphics[width=20pc]{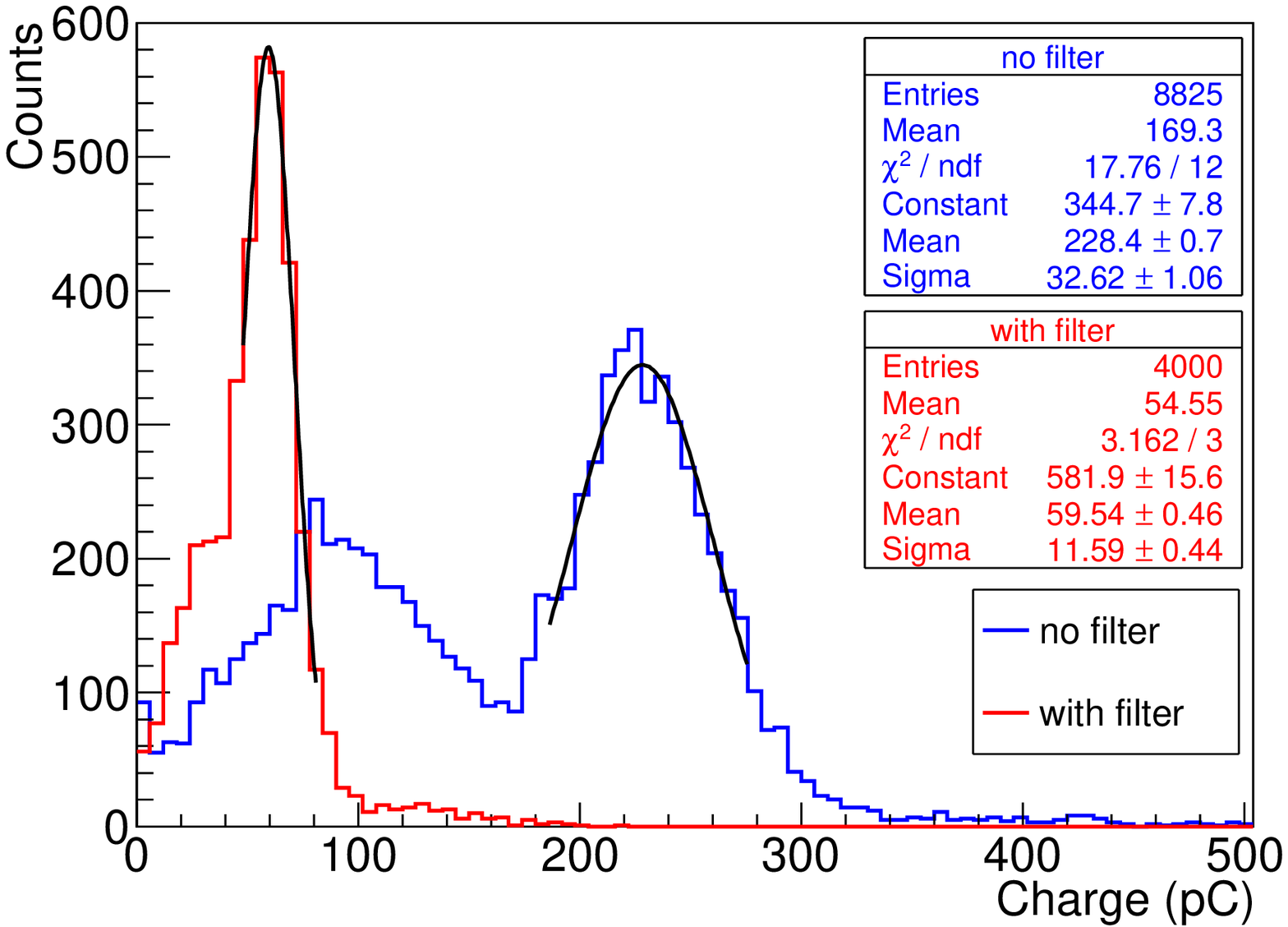}
\caption{\label{spectra_scope}Spectra from the BaF$_{2}$ crystal irradiated with the $^{22}$Na gamma source without and with a filter. The data are collected with the HDO4104 digitizing oscilloscope.}
\end{minipage}
\end{figure}

Spectra from the BaF$_{2}$ crystal irradiated with the $^{22}$Na gamma source taken without and with a filter are presented in Figure \ref{spectra_QDC}. Data were collected by the 2249W QDC with the gate width of 2 $\mu$s. The results demonstrate that the total signal from the BaF$_{2}$ crystal was suppressed by a factor of 3.5 when the data are taken with a filter.

Similar spectra were taken with the HDO4104 digitizing oscilloscope. The results are shown in Figure \ref{spectra_scope} for both cases with no filter and with a filter. In that case we got the suppression of the total signal by a factor of 3.8. The difference between two methods could be explained by inaccurate calibration.

\begin{figure}[h]
\centering
\includegraphics[width=0.45\textwidth]{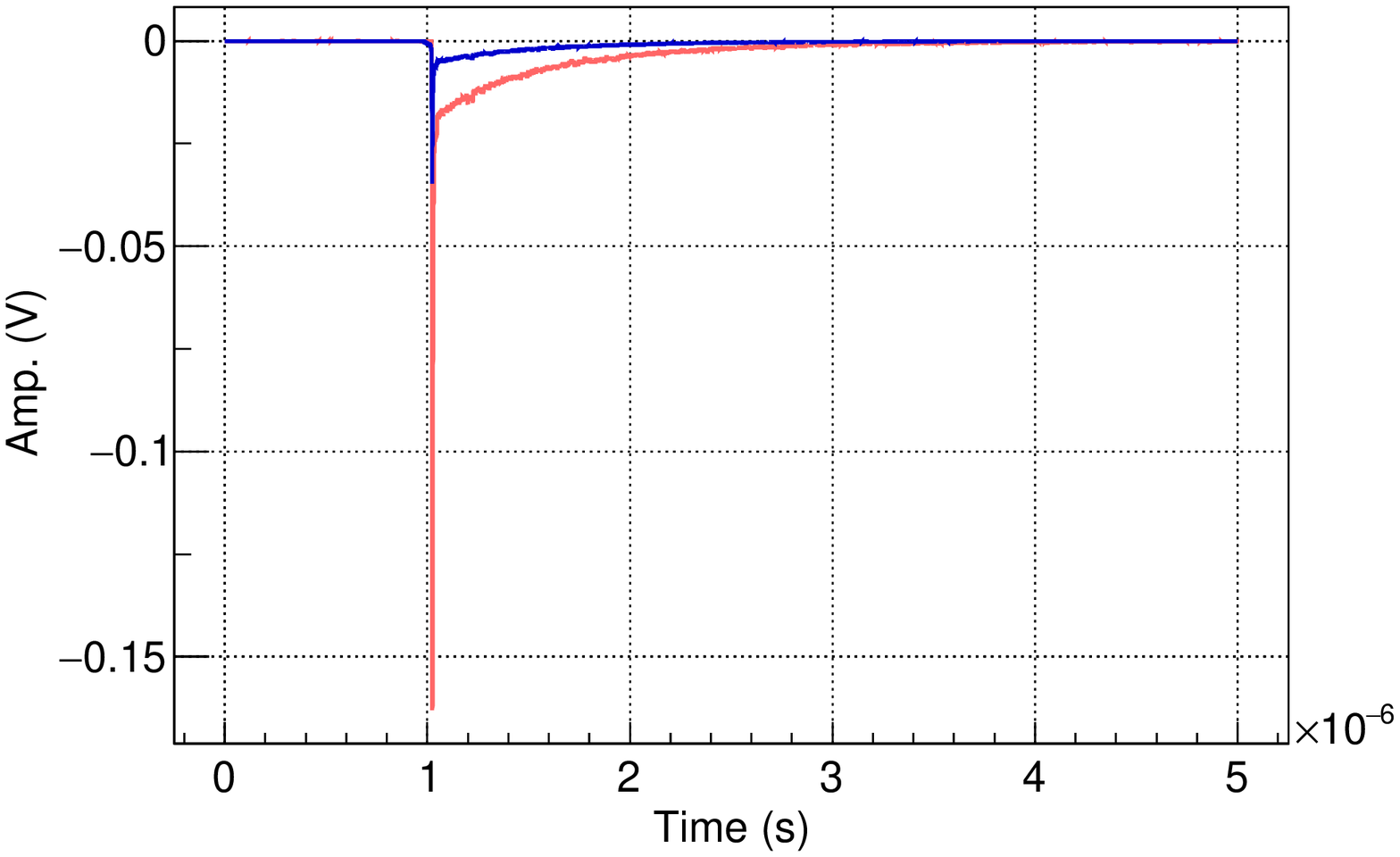}
\includegraphics[width=0.45\textwidth]{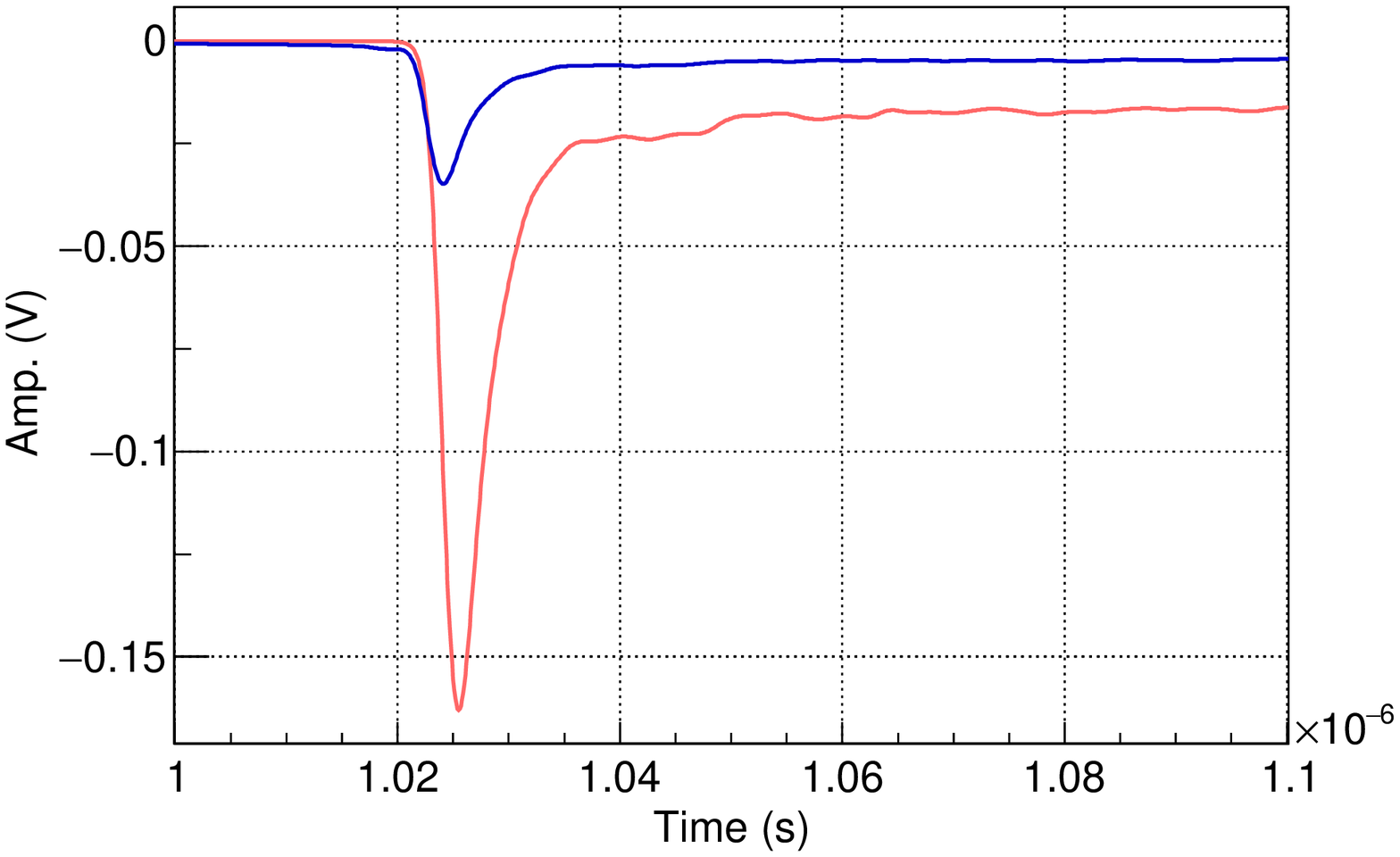}
\caption{Averaged signals from the BaF$_{2}$ crystal with and without a filter. Left: full 6 $\mu$s range; right: details of the fast components.}
\label{averaged_WF}
\end{figure}

In order to study the light output and decay kinetics
 we selected 1100 events from the centers of both full absorption peaks of the spectra taken with the HDO4104 digitizing oscilloscope (see Figure \ref{spectra_scope}). Averaged signals of these two 1100-event sets are shown in Figure \ref{averaged_WF}: the left frame presents full scale signals within 5 $\mu$s, and the right frame gives details of the fast components. Right frame confirms that fast component is suppressed about by factor 4 as well.

\begin{figure}[h]
\begin{minipage}{18pc}
\includegraphics[width=18pc]{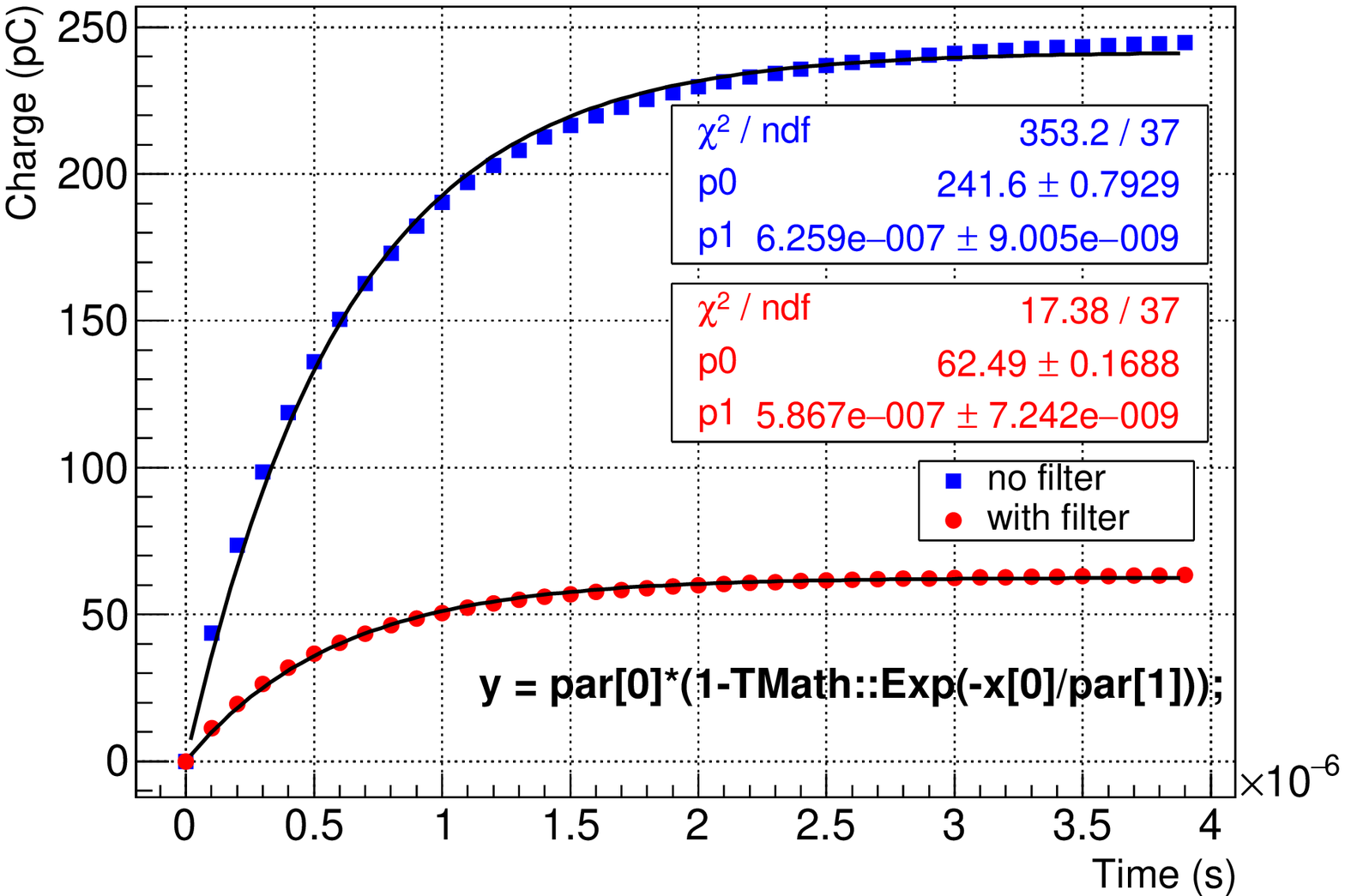}
\caption{\label{averaged_charge}Charge collected from the BaF$_{2}$ crystal without and with a filter as a function of integration time.}
\end{minipage}\hspace{2pc}%
\begin{minipage}{18pc}
\includegraphics[width=18pc]{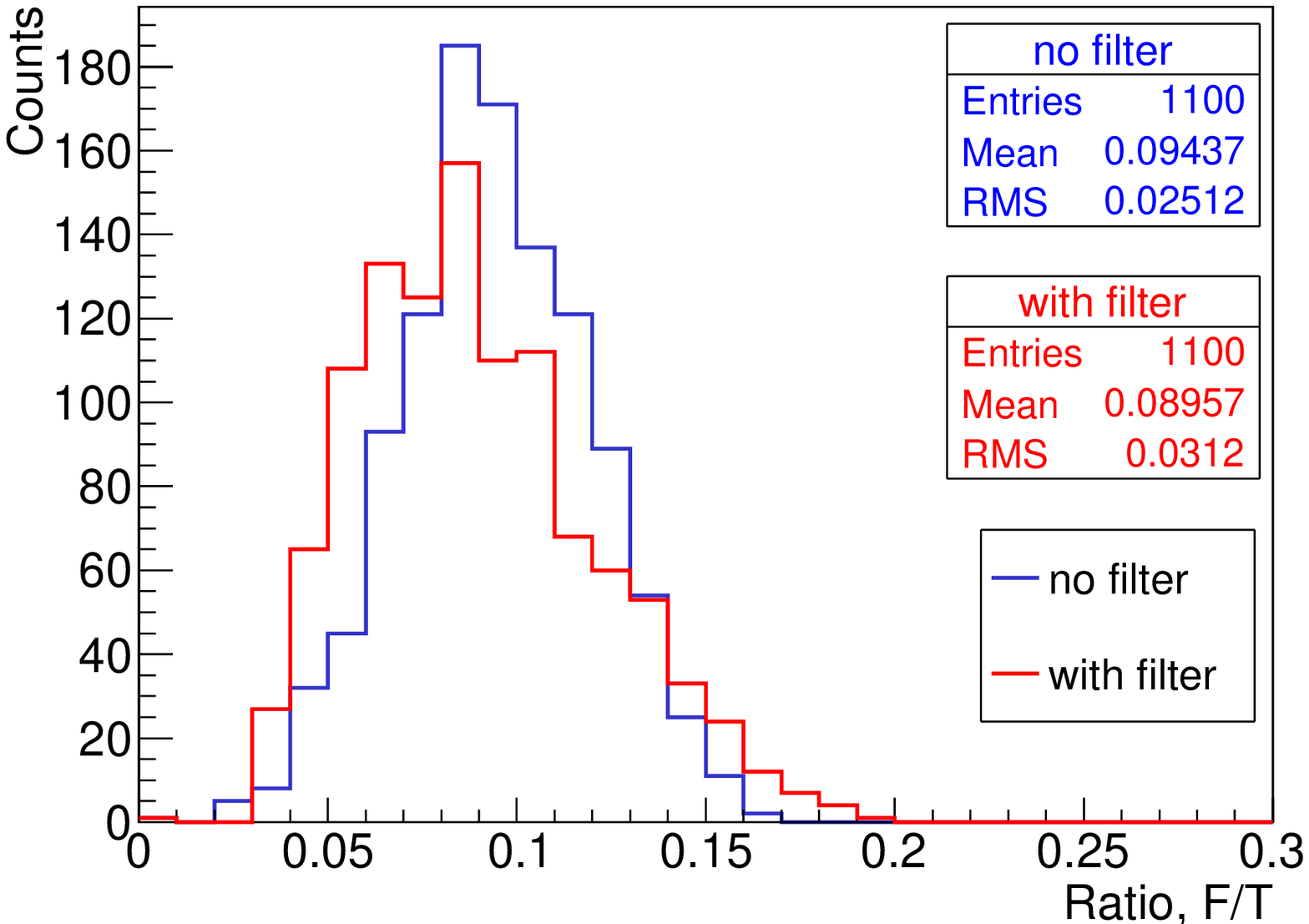}
\caption{\label{FT_ratio}Fast/Total ratio for signals from the BaF$_{2}$ crystal without and with a filter.}
\end{minipage}
\end{figure}

Averaged signals were integrated within the full scale in order to get the total charge from the crystal. The results are depicted in Figure \ref{averaged_charge}. One can see that total signals with a filter are suppressed by a factor of about 4 compared with those without a filter.

Ratios of Fast/Total signals for 1100 selected events taken both with  and without a filter are presented in Figure \ref{FT_ratio}. The fast component was defined as a charge collected within the first 20 ns, while the total signal corresponds to the charge collected within 2 $\mu$s. Almost equal Fast/Total ratios for signals, taken both with and without a filter show that in our case the fast component is suppressed at the same level as the slow component and confirms the results seen in Figures \ref{Filter_transmit} and \ref{averaged_WF}.

\section{Conclusions}

Thin multilayer interference filters made of up to 200 layers of  rare earth oxides have been produced and tested with the BaF$_{2}$ crystal. To our knowledge, this kind of filters were produced and tested for the first time.

Such thin multilayer filters can suppress luminescence in the range from 250 nm to 400 nm and could be used for suppression of the slow component of BaF$_{2}$ crystals.

Tests of the filters made by spraying thin layers of rare earth oxides on a quartz glass substrate demonstrated that they suppress the total signals from the BaF$_{2}$ crystal by a factor of 4. However, the results show that the fast component is suppressed by these developed filters as well. It is obvious that more research is required to improve the quality of the multilayer filters.


\section*{References}
\begin{thebibliography}{99}
\bibitem{Mu2e_TDR} Mu2e Collaboration (L. Bartoszek {\it et al.}), Mu2e Technical Design Report, arXiv:1501.05241.
\bibitem{Mu2e-II} Mu2e Collaboration (F. Abusalma {\it et al.}), Expression of Interest for evolution of the Mu2e experiment, arXiv:1802.02599.
\bibitem{BaF2_suppress} B.P. Sobolev {\it et al.}, Suppression of BaF2 slow component of X-ray luminescence in non-stechiometric Ba$_{0.9}$R$_{0.1}$F$_{2.1}$ crystals (R=Rare Earth element,) {\it Proceedings of The Material Research Society: Scintillator and Phosphor Materials}, p.277, (M. J. Weber, ed.), San Francisco, CA, 1994.
\bibitem{Hitlin} D.G. Hitlin, J.H.Kim, J.Trevor, M.Hoenk, J.Hennessy, A.Jewell, R.Farrell and M. McClish, An APD for the efficient detection of the fast scintillation
component of BaF$_{2}$, 2016, {\it Nucl Instrum Meth A} {\bf 824} 119.
\end{thebibliography}
\end{document}